\documentclass[preprint2]{aastex}

\slugcomment{}

\shorttitle{Solar radiation belts}
\shortauthors{Hudson et al.}

\begin{document}


\title{Coronal radiation belts}


\author{H.~S. Hudson}
\affil{Space Sciences Laboratory, University of California,
    Berkeley, CA 94720}

\author{A.~L. MacKinnon}
\affil{University of Glasgow, UK}

\author{M.~L. DeRosa}
\affil{Lockheed Martin Solar and Astrophysics Laboratory, 3251 Hanover St B/252, Palo Alto, CA 94304}

\author{S.~F.~N. Frewen}
\affil{Space Sciences Laboratory, University of California,
    Berkeley, CA 94720}




\begin{abstract}
The magnetic field of the solar corona has a large-scale dipole character, which maps into the bipolar field in the solar wind.
Using standard representations of the coronal field, we show that high-energy ions can be trapped stably in these large-scale closed fields.
The drift shells that describe the conservation of the third adiabatic invariant may have complicated geometries.
Particles trapped in these zones would resemble the Van Allen belts and could have detectable consequences.
We discuss potential sources of trapped particles.
\end{abstract}





\section{Introduction}

The large-scale magnetic field of the solar corona may trap high-energy particles for long periods of time, according to early suggestions \citep[e.g.,][]{1969sfsr.conf..356E}.
The original impetus for considering coronal trapped particles came from the presence of long timescales (hours to days) and the appearance of diffusive transport of solar energetic particles (SEPs, then often called ``solar cosmic rays'').
Diffusive transport suggested coronal trapping and ``leakage' into the heliosphere.
More recently it has become clear that SEPs can be accelerated in heliospheric shock fronts 
\citep{1988JGR....93.9555C,1999SSRv...90..413R}, and that the long timescale for injection into the heliosphere could be associated with the propagation time of the shock.
Because the shock disturbance takes hours to days en route to 1~A.U., a leaky coronal trap now seems unnecessary to explain SEPs.
It should be noted, however, that observations of the 2.223~MeV $\gamma$-ray line due to neutron
capture have already suggested trapping for many hours even in active-region fields \citep[e.g.][]{2000SSRv...93..581R}.

There are several other implications of coronal particle storage, to the extent that plasma instabilities permit it \citep[e.g.,][]{1976ApJ...208..595W}.
In particular, as noted by Elliot and others, the particle storage may involve a substantial
energy density and perpendicular pressure.
The particles could therefore have effects on the structure of the coronal plasma.
However the ion component, not coupling well to detectable electromagnetic radiation, would mainly be identifiable via its secondary effects.

The geometry of the large-scale coronal magnetic field, discussed below in Section~\ref{sec:field},
strongly suggests a basically dipole character at solar minimum and in the high corona.
If this were true, then we would expect the existence of ``forbidden zones'' in St{\o}rmer's sense, with the
possibility of long-term trapping of particles resulting from cosmic or solar sources.
In this Letter we revisit the geometry of coronal particle trapping and demonstrate that it can be complete, in the sense that all three of the adiabatic invariants of charged-particle motion 
\citep[e.g.][]{northrop.book.....B} may be conserved for times long enough for particles to circulate completely around the Sun and thus potentially complete closed trapping shells.
Section~\ref{sec:field} describes the methods used to establish this result, which is based on analysis
and modeling.
The detailed modeling uses the \cite{2003SoPh..212..165S} version of the ``potential field source surface'' (PFSS) approach to the solar large-scale fields \citep{1969SoPh....9..131A,1969SoPh....6..442S}.
This type of model uses the routine Zeeman-splitting observations of the line-of-sight photospheric magnetic field as an inner boundary condition, and assumes a radial field outside a fixed ``source surface'' at (in this case) 2.5~R$_\odot$.
In between the photosphere and the source surface, the model simply uses a potential expansion of the field; the body currents known to flow in the corona are replaced by currents outside the source surface.
This approach appears to work reasonably well for a large-scale structure.
In Section~\ref{sec:particles}, we summarize the experiments we have made using PFSS fields.
Even in these relatively complicated fields we find trapping orbits that allow particles to circulate all the way round the Sun, potentially conserving the third adiabatic invariant \citep[e.g.][]{northrop.book.....B}.

\section{The coronal magnetic field}\label{sec:field}

Eclipse pictures show the solar-minimum corona to have a strongly dipolar configuration, especially at solar minimum.
Hale's discovery of solar magnetism allowed an explanation of this appearance by analogy with the
Earth's field, which comes from sources deep in the interior and has a true dipole nature.
The coronal field has a fundamental difference in that at low altitudes small-scale fields dominate; these fields originate in the magnetic active regions or in the ``network'' structure \citep[e.g.,][]{1987ARA&A..25...83Z,1993PhDT.......225H}.
The dipolar character only appears at higher altitudes in the corona, where the intense small-scale fields of solar active regions and the network do not extend.
The dipole structure thus appears only on the largest scales, and is mainly visible in eclipse images at solar minimum.
The observed polarity structure of the magnetic field in the solar wind normally has a simple ``sector'' structure \citep{1965JGR....70.5793W}.
It is thus bipolar although highly non-potential in character since the field is largely radial.

Our detailed knowledge of the coronal field remains meager, with almost negligible direct measurement because of the difficulty of making precise polarization measurements in the corona \citep[e.g.][]{2000ApJ...541L..83L}.
There are several approaches to extrapolating the photospheric field into the corona instead.
These can be compared with in situ measurements of the heliospheric field, but have the
drawback that the treatment of currents penetrating the photosphere and flowing in the corona
is ambiguous even in the ``nonlinear force-free field" (NLFFF) extrapolations \citep[e.g.,][]{2006SoPh..235..161S}.
Full~MHD simulations of coronal structure \citep[e.g.,][]{2007ASPC..370..299M} may improve our detailed understanding.
In the meanwhile, the Schrijver~\&~DeRosa PFSS computational structure is very convenient, because 
it runs as a software package in SolarSoft \citep{1998SoPh..182..497F} and the database gives 6-hour time resolution since~1996, based on \textit{Solar and Heliospheric Observatory}/Michelson Doppler Imager~\citep{1995SoPh..162..129S}.
We can use this model to explore particle trapping in the coronal field, but note that it has physical limitations.
Indeed, neither PFSS nor more any of the more sophisticated coronal field models have succeeded in reliably detecting the obvious restructurings due to flares and coronal mass ejections (CMEs), which have clear photospheric counterparts \citep[cf.][]{2005ApJ...635..647S}.

\section{Particle propagation}\label{sec:particles}

In a low-order analytic model of the field such as a dipole or a Mead field\footnote{The Mead fields  \citep{1964JGR....69.1181M} are distorted dipoles with a full analytical description, originally intended as a representation of geomagnetic field.
We use these to test our codes.}, the length scales of the field greatly exceed the particle
gyroradii, even for the heaviest ions we consider.
The same should be true of the~PFSS representation or other extrapolations of the photospheric field with limited spatial sampling.
We therefore solve the relativistic guiding-center drift equations of motion as given by \cite{northrop.book.....B} or \cite{2004pspi.book.....P}, using spherical polar coordinates.
From the point of view of particle trapping in the corona, the question is whether or not particles can
readily move via gradient or curvature drifts from closed to open field lines, and hence ``escape,''
and whether their mirror points move low enough for precipitation.
Particles that can survive either of these threats or collisional energy decay then can drift completely around the Sun, potentially conserving the third adiabatic invariant and defining a drift shell.

\begin{figure}
\plotone{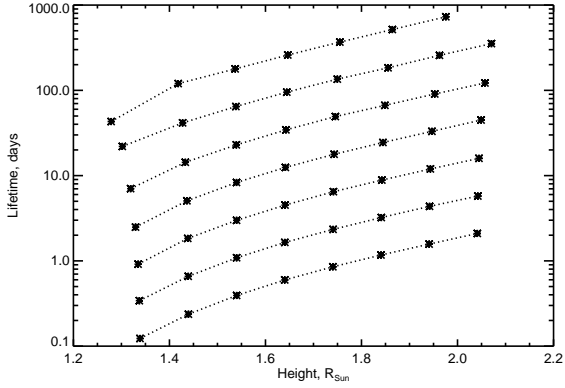}
\caption{Lifetimes ($E/\dot{E}$) of energetic protons in a representative PFSS field mode( 2006 Sep. 30)l, for different  injection altitudes at  ($\theta, \phi$)~=~(6.02, 1.04).
The sets of points correspond to proton energies of 4, 8, 16... 256~MeV.}
\label{fig:lives}
\end{figure}

Our modeling here aims at defining the morphology of the particle trapping, rather than providing
quantitative results.
We follow particle motions with an RK4 (Runge-Kutta)  integrator \citep{1986nras.book.....P} as
implemented in IDL, for convenience, and typically use a heavy test particle (mass number 207, for example) to exaggerate the cross-field drifts.
The collisional energy losses in the coronal plasma are estimated from \cite{2002NIMPB.187..285W}, who provide corrections to the basic formula $-dE/dx \propto Z^2 n_e L/v^2$, where $Z$ is the projectile charge, $v$ is its velocity, $n_e$ is the electron density, and $L$ is the ``stopping logarithm.'' 
For an ionized medium we  take the ionization potential to be $\hbar \omega_p$, where $\hbar$~is Planck's constant, and $\omega_p$~is the plasma frequency.
We adopt the Van de Hulst-Newkirk density model \citep{1961ApJ...133..983N}, and Figure~\ref{fig:lives} shows representative lifetimes for 1-100~MeV protons.
The coronal magnetic field is not really understood well enough to justify detailed calculations; the main concern here is that the PFSS approach, though convenient, systematically misrepresents it both in the low corona where field-aligned currents are strong \citep[e.g.,][]{2008ApJ...675.1637S} and also in the high corona near the ad hoc source surface at 2.5~R$_{\odot}$.
Thus we would trust our numerical experiments on trapping mostly in the middle corona, at heights of 1.5-2.0~R$_{\odot}$.
It would clearly be possible to do similar experiments with more complete models, such as those based on the MHD equations \citep[e.g.][]{2007ASPC..370..299M}.

\section{Existence and stability of trapping zones}

\begin{figure}
\plotone{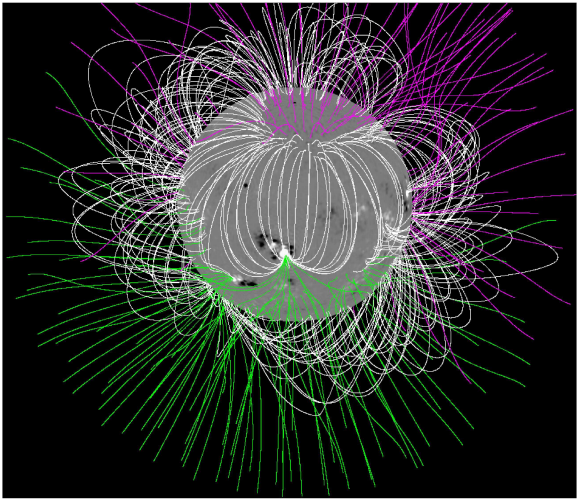}
\caption{PFSS field for 2006 September 30, 12:04 UT. Open field lines are in color, closed in white.}
\label{fig:pfss}
\end{figure}

After testing our code on analytical field models (dipole and Mead), we have explored parameter space by injecting particles with specified energies and pitch angles into PFSS coronal models.
The initial experiments were with the PFSS field realization of 2006 September 30 12:04 UT, selected just for its simple appearance (see Figure~\ref{fig:pfss}).
The easily seen non-dipolar structure shows that this epoch is not the very simplest that could have been chosen, and indeed four GOES X-class flares occurred subsequently in 2006 December.

\begin{figure}
\plotone{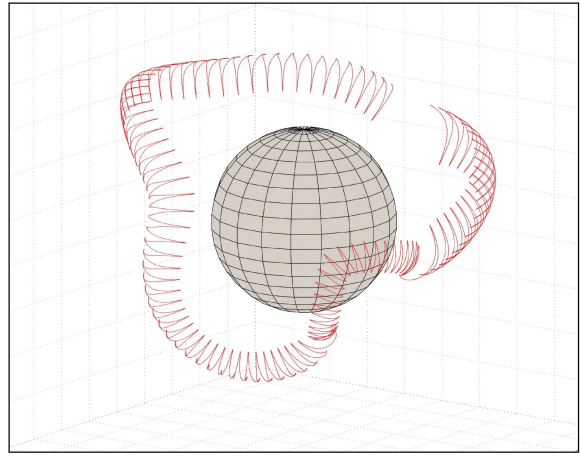}
\caption{Trajectory of an individual test particle (Bi$^{+83}$ to exaggerate the non-adiabatic effects) in the PFSS field of Figure~\ref{fig:pfss}.
The initial energy is 150~MeV, pitch angle 90$^\circ$, and coordinates ($\theta, \phi$, R) = (6.02, 1.04, and 2.0~R$_\odot$).}
\label{fig:shane}
\end{figure}

Figure~\ref{fig:shane} shows the trajectory of a single test particle injected at a chosen point (near the middle of the estimated trapping zone in this case).
We used an unusual ion (Bi$^{+83}$) to exaggerate the losses.
The bouncing motion results from the fact that the injection point (height 2.0~R$_{\odot}$) was \ at the apex of the field line intersecting the injection point, rather than its minimum value of $|\textbf{B}|$.
This particle completed its full shell motion in about 6~hr and could be followed for more than one circulation.
It could be argued that the PFSS model is not realistic enough, simply because the field reconstruction is limited to a small number of spherical harmonics, and hence cannot show steep gradients and large non-adiabatic effects.
On the other hand we are using a test ion with a Larmor radius orders of magnitude too large, and we are experimenting around 2~R$_\odot$ where the field presumably cannot be stressed strongly on small scales.
We thus feel that this approximation is realistic enough to establish the principle that trapping zones exist in the corona.

\begin{figure}
\plotone{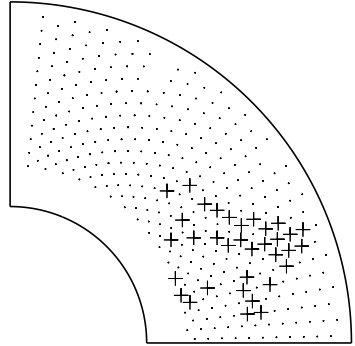}
\caption{Sector of the meridional plane (60$^\circ \times$~1.5~R$_{\odot}$) at heliolongitude 6.02~rad for the same date as the simulations
shown above.
The plus signs mark the long-lived orbits for particles injected near the tops of the field lines
intersecting the grid of points, with zero initial parallel velocity.
}
\label{fig:contour}
\end{figure}

How large is the volume of phase space showing this property?
Figure~\ref{fig:contour} shows the results for test particles in a half of the meridional plane ($\theta$, $R$-space) at the injection longitude (6.02~rad) for the particle of Figure~\ref{fig:shane}.
Particles injected at zero pitch angle (but not necessarily zero parallel velocity) have trapped orbits, and the remainder either start on open field lines or drift onto them, with a few particles at the lower boundary precipitating.
We conclude from this and other tests that drift shells (stable trapping) occupy a substantial fraction of the coronal phase space for field lines closing above 1.4~R$_{\odot}$.
The drift-shell regions may not be contiguous.

\section{Sources of particles}

In the Earth's radiation belts, the ``Cosmic Ray Albedo Neutron Decay'' (CRAND) mechanism provides a basic guarantee that some particles can materialize in the Van Allen trapping zones.
The same is true for the Sun, except that the cosmic-ray flux is diminished by heliospheric modulation \citep[e.g.][]{1991ApJ...382..652S}.
In addition the mean atomic number of the solar atmosphere is lower, around 1.3 versus 7.2 for the Earth's atmosphere.
This means more $(p, p)$ primary interactions for the primary cosmic rays, which produce fewer neutrons than do spallation reactions on higher-$Z$ nuclei.

There are good reasons to think that trapping zones in the coronal magnetic field will actually contain particles.
Large-scale shock waves propagate through the region as a result of flares and/or CMEs.
Shock acceleration \citep[e.g.][]{1999SSRv...90..413R} is known to be an important factor in the production of SEPs.
We cannot be quantitative now about the possible contributions of this mechanism to the trapping zones, especially since the CME closely involved with causing the shock, and the particle acceleration, itself may disrupt the geometry of the corona.
Charge exchange at MeV energies recently has been observed via flare-associate energetic neutral atoms observed by STEREO \citep{2009ApJ...693L..11M}.
Although only a single event (2006 December 5) has thus far been reported, this clear observation strongly suggests that  SEP-associated flares will populate distant regions of the corona via neutralization and reionization interactions.
Similarly, the copious production of neutrons by flares, as evidenced by the 2.223~MeV $\gamma$-ray emission and by their direct detection at Earth, guarantees a source of particles at some level.
Neutron-decay particles have also been detected directly in the heliosphere \citep{1983ApJ...274..875E}. 
Other mechanisms operate in the Earth's radiation belts, and via pitch-angle diffusion the trapped particles can actually gain energy as they migrate to regions of stronger field.

\section{Conclusions}

We have made basic models of high-energy particle propagation in the solar corona, using PFSS extrapolations of the magnetic features observed at the photospheric level.
The PFSS models, while not perfect, have a reasonable track record for establishing connectivity on large scales. 
Our calculations show that trapping zones exist in the corona, just as they do in the Earth's field.
In these zones the main ``escape route'' for a trapped particle will be its collisional losses at lower
energies.

While it is not possible now to study these particles in situ, as one can do near the Earth, their presence may have indirect signatures.
These could include $\gamma$-ray, X-ray, or radio emissions, and the particles could also have dynamical consequences for the structure of the plasma if numerous enough.
Direct remote sensing of the ion component is almost nonexistent, but  high-energy electrons could possibly be detectable.
Various mechanisms exist \citep[e.g.][]{2006ApJ...652L..65M,2008ApJ...675..846K}, and both EGRET \citep{2008A&A...480..847O} as-yet-unpublished \textit{Fermi} results show a distinct source of high-energy $\gamma$~rays from the solar direction.
The particle environment of the corona itself may play a role in understanding this source and its relationship to solar-system background sources for \textit{Fermi} observations.

\acknowledgments
This work was supported by NASA under grant NAG5-12878 (HSH and SF).
Solar physics at Glasgow (ALM) is supported by an SFTC Rolling Grant, and ALM thanks his Berkeley
colleagues for hospitality during a sabbatical visit.


\clearpage

\end{document}